\documentstyle[aps,epsf]{revtex}

\def\inseps#1#2{\def\epsfsize##1##2{#2##1}
\centerline{\epsfbox{#1}}}

\begin{document}

\title{Dissipative particle dynamics: the equilibrium for finite time steps.}
\author{C. A. Marsh and J. M. Yeomans}
\address{Theoretical Physics, Oxford University, 1 Keble Rd. Oxford OX1 3NP, UK}
\date{\today}
\maketitle
\begin{abstract}

Dissipative particle dynamics (DPD) is a relatively new technique which has
proved successful in the simulation of complex fluids. We caution that
for the equilibrium achieved by the DPD simulation of a simple fluid
the temperature depends strongly on the time step. An analytic
expression for the dependence is obtained and shown to agree well with
simulation results.

\end{abstract}
\vskip 50pt
PACS numbers: 02.70.Ns; 47.11.+j; 05.70.Ce
\vskip 100pt
\newpage

Modelling the rheological behaviour of complex fluids using standard
numerical techniques is very difficult and often impossible. Molecular
dynamics simulations which can faithfully represent the microscopic
nature of the fluid require intesive computational power to reach time
scales over which hydrodynamic
effects are operative. Macroscopic approaches, based on the
Navier-Stokes equations, must include phenomenological constitutive
relations which are difficult to verify.

In recent years a new class of techniques for hydrodynamic simulation
have been developed. These include lattice-gas cellular 
automata \onlinecite{RZ94} ,
lattice Boltzmann simulations \onlinecite{BSV92} , 
and dissipative particle dynamics \onlinecite{HK92} . 
In
some sense these methods may be termed mesoscopic because the fluid is
modelled on a length scale that allows input of the relevant physics
but not the details of the interatomic interactions. For example, a
long polymer may be represented by a few particles connected by
springs in the spirit of the Rouse-Zimm model \onlinecite{SHM95} . 
Hydrodynamics then
follows from the constraints of conservation of local mass
and momentum. Although application of these techniques to complex
fluids is in its infancy promising results have been obtained for
several systems including microemulsions \onlinecite{BCE96} ,
colloidal suspensions \onlinecite{L94} , and multiphase flow in 
porous media \onlinecite{GR93} .

One of the most flexible but least explored of the approaches is
dissipative particle dynamics (DPD) \onlinecite{HK92} . 
A set of particles, each of which
may be interpreted as representing a mesoscopic region of fluid, move
in continuous space and discrete time. Each particle is subjected to
Brownian noise, and a friction force acts between them in such a way
that mass and momentum are conserved. As pointed out by Espa\~nol and
Warren \onlinecite{EW95} , DPD is an extension of the Langevin 
equation of Brownian
motion to a system which conserves momentum as well as mass and hence
obeys hydrodynamics.

DPD has been used to simulate phase separation in a binary fluid by
designating the particles to be of two different types and adding a
conservative repulsive force between unlike
particles \onlinecite{CN96} . 
It has been
applied to dilute solutions of polymers by choosing a few of the
particles to represent sections of the polymer chain and adding
springs between them \onlinecite{SHM95} . 
A third application is to the observance of shear
thinning in particulate suspensions \onlinecite{BCL96} . 
In each case the simulations have
been very successful in reproducing the expected behaviour.

The success of the method is surprising given its seemingly ad hoc
nature. Much remains to be understood about the theoretical
justification for the approach. In an important first step Espa\~nol
and Warren \onlinecite{EW95} , showed that, given the correct 
relation between the forms of
the random and dissipative forces, the system relaxes to a Gibbs'
distribution characterised by a temperature related to the noise
amplitude via a fluctuation-dissipation theorem. However their results
hold only in the limit that the time step becomes infinitesimal. This
is a severe limitation because the power of DPD relies on its ability
to take large time steps in order to probe long time scales.

Therefore our aim in this letter is to study the equilibrium of
the system for a general time step.  Away from the limit, $\Delta t
\rightarrow 0$, it is not obvious that the equilibrium distribution is the
Gibbs distribution.  However, it is expected that, close to this
limit, it will be a good approximation.  Numerical simulations verify
this.  We demonstrate that, under this assumption, the distribution
would remain unchanged under the full finite $\Delta t$ evolution and derive
the dependence of the corresponding temperature T
on the time step $\Delta t$ and, it transpires, the density n. 
The dependence is large. For example, for $n=0.2$, $T(\Delta t=0.25)
\sim 2 \times  T(\Delta t =0)$.  

We first describe the DPD algorithm in more detail. Then we derive an 
expression for
the temperature in terms of the parameters involved in the
updating step. The formula is compared to simulation results.

Consider a set of particles $i$ of equal mass $m$ at positions
$\vec{r}_i$ with momentum $\vec{p}_i$. The system is updated according
to the algorithm \onlinecite{HK92}
\begin{equation} \label{eqn1}
r_i^{\alpha}(t+\Delta t) - r_i^{\alpha} (t) \equiv
\Delta r^{\alpha}_{i} = \frac{p^{\alpha}_{i}}{m} \Delta t
\end{equation}
\begin{equation} \label{eqn2}
p_i^{\alpha}(t+\Delta t) - p_i^{\alpha} (t) \equiv
\Delta p^{\alpha}_{i} = \sum_{j \neq i} \left\{
-\gamma w_{D}(r_{ij}) ({e_{ij}^{\beta}}{v_{ij}^{\beta}}) \Delta t + \sigma
w_{R}(r_{ij}) \xi_{ij} (\Delta t)^{\frac{1}{2}} \right\} e_{ij}^{\alpha}
\end{equation}
where superscripts $\alpha, \beta, \ldots$ are used to represent
Cartesian components of a vector and the usual summation convention is
assumed for the Cartesian labels. Subscripts $i,j, \ldots$ distinguish
different particles. $\vec{e}_{ij}$ is the unit vector between the
$i^{th}$ and $j^{th}$ particles, $r_{ij}$ their separation and
$\vec{v}_{ij} \equiv \vec{v}_i -\vec{v}_j$ their relative
velocity. $\gamma$ is the strength of the dissipative force and
$\sigma$ the strength of the random force and $w_D(r_{ij})$ and
$w_R(r_{ij})$ are radial weighting functions for each. The random
variables $\xi_{ij}$ obey
\begin{eqnarray} \label{eqn3}
\xi_{ij}=-\xi_{ji}, \;\;\;&
\overline{ \xi_{ij}(t) } = 0,\;\;\; &
\overline{ \xi_{ij}(t)\xi_{kl}(t^{\prime}) } =
(\delta_{ik}\delta_{jl} + \delta_{il}\delta_{jk})\delta_{tt'}
\end{eqnarray}
where a bar represents an average over the ensemble of all $\xi_{ij}$.
The evolution algorithm can be written more conveniently as
\begin{equation} \label{eqn4}
\Delta p^{\alpha}_{i}(t) = 
\sum_{j \beta} L^{\alpha \beta}_{ij}
p^{\beta}_{j} + \sum_{j} M^{\alpha}_{ij} \xi_{ij}
\end{equation}
where
\begin{eqnarray} \label{eqn5}
L^{\alpha \beta}_{ij} &=& - \delta_{ij} \frac{\gamma}{m} \sum_{l} w_{D}(r_{il})
e_{il}^{\alpha} e_{il}^{\beta} \Delta t +
\left[ 1- \delta_{ij} \right] \frac{\gamma}{m} w_{D}(r_{ij})
e_{ij}^{\alpha} e_{ij}^{\beta} \Delta t, \\
M^{\alpha}_{ij} &=& \sigma w_{R}(r_{ij}) e_{ij}^{\alpha}
(\Delta t)^{\frac{1}{2}} \left[ 1-\delta_{ij} \right]. \label{eqn6}
\end{eqnarray}

Let the N-particle distribution function of the system be
$f^{(6N)}(\Gamma_{p,r})$,
where $\Gamma_{p,r}$ represents all the momenta and position variables
of N particles. The one-particle distribution function can then be
defined as \onlinecite{RD77} 
\begin{equation} \label{eqn7}
f^{(1)}(\vec{r}, \vec{p}, t) = \sum_{i} \int d\Gamma_{p,r} 
\delta (\vec{r}-\vec{r_{i}}(t)) \delta (\vec{p}-\vec{p_{i}}(t))
f^{(6N)}(\Gamma).
\end{equation}

For $\Delta t \rightarrow 0$, it has been
demonstrated \onlinecite{EW95}  
that the equilibrium distribution function in the
absence of a conservative force is the Gibbs distribution
\begin{equation} \label{eqn8}
f^{(6N)}_{eq} = \frac{1}{Z} \exp \left\{
- \frac{1}{k_{B}T} \sum_{i} \frac{\vec{p_{i}}^{2}}{2m}
\right\}.
\end{equation}
We shall assume that this distribution
function also provides an  equilibrium solution for $\Delta t$ 
finite and derive the
constraints on the system for this to be consistent.

The change in the one-particle distribution function 
between times $t$ and $t + \Delta t$ is given by
\begin{equation} \label{eqn9}
\Delta f^{(1)} = \sum_{i} \int d\Gamma_{p,r} \left\{
\delta (\vec{r}-\vec{r_{i}}-\Delta \vec{r_{i}}) 
\delta (\vec{p}-\vec{p_{i}}-\Delta \vec{p_{i}})
- \delta (\vec{r}-\vec{r_{i}}) \delta (\vec{p}-\vec{p_{i}}) \right\}
f^{(6N)}_{eq}.
\end{equation}
Expanding the integrand
\begin{eqnarray} \label{eqn10}
\Delta f^{(1)} &=&\sum_{i} \int d\Gamma_{p,r} \left\{
-\frac{p_{i}^{\alpha}(\Delta t)}{m}
\frac{\partial}{\partial r^{\alpha}} - \Delta p_{i}^{\alpha}
\frac{\partial}{\partial p^{\alpha}} + 
\frac{p_{i}^{\alpha} p_{i}^{\beta}}{m^2}\frac{(\Delta t)^{2}}{2}
\frac{\partial^{2}}{\partial r^{\alpha} \partial r^{\beta}} \right.
\nonumber \\
&&\left. +\frac{\Delta p_{i}^{\alpha} \Delta p_{i}^{\beta}}{2}
\frac{\partial^{2}}{\partial p^{\alpha} \partial p^{\beta}}
+\frac{p_{i}^{\alpha} (\Delta t)}{m} \Delta p_{i}^{\beta}
\frac{\partial^{2}}{\partial r^{\alpha} \partial p^{\beta}}
+ \ldots  \right\}f_{eq}^{6N} \delta (\vec{r}-\vec{r_{i}})
\delta (\vec{p}-\vec{p_{i}}).
\end{eqnarray}
Defining the  averages
\begin{eqnarray} \label{eqn11}
\langle A \rangle_{p,r} &=&
\sum_{i} \int d\Gamma_{p,r} f^{(6N)}_{eq} \delta
(\vec{r}-\vec{r_{i}}) \delta (\vec{p}-\vec{p_{i}}) A(\Gamma_{p,r}) \\
\langle A \rangle_{r} &=&
\sum_{i} \int d\Gamma_{r} f^{(6N)}_{eq} \delta
(\vec{r}-\vec{r_{i}})) A(\Gamma_{r}) 
\end{eqnarray}
gives
\begin{eqnarray} \label{eqn13}
\Delta f^{(1)} &=& -\frac{(\Delta t)}{m} \frac{\partial}{\partial r^{\alpha}}
\langle p_{i}^{\alpha} \rangle_{p,r}
-\frac{\partial}{\partial p^{\alpha}} \langle \Delta p_{i}^{\alpha}
\rangle_{p,r}
+\frac{(\Delta t)^{2}}{2m^{2}} \frac{\partial^{2}}{\partial r^{\alpha}
\partial r^{\beta}} \langle p_{i}^{\alpha} p_{i}^\beta \rangle_{p,r}
\nonumber \\
&+& \frac{1}{2} \frac{\partial^{2}}{\partial p^{\alpha} \partial p^{\beta}}
\langle \Delta p_{i}^{\alpha} \Delta p_{i}^{\beta} \rangle_{p,r}
+\frac{(\Delta t)}{m} \frac{\partial^{2}}{\partial r^{\alpha} \partial
p^{\beta}} \langle p_{i}^{\alpha} \Delta p_{i}^{\beta} \rangle_{p,r} + \ldots.
\end{eqnarray}

We shall follow the evolution of $f^{(1)}$ under the finite time step
algorithm.  The easiest way to do this is to calculate the changes in the
moments of this distribution function.
Firstly consider
\begin{equation} \label{eqn14}
\Delta \int d\vec{p} \int d\vec{r} f^{(1)} \vec{p}= 
\int d\vec{p} \int d\vec{r} \Delta f^{(1)} \vec{p}.
\end{equation}
Upon substitution of equation (\ref{eqn13}), integration by 
parts removes all terms
except for that involving only one momentum derivative. This gives
\begin{equation} \label{eqn15}
\Delta \int d\vec{p} \int d\vec{r} f^{(1)} \vec{p} =
\int d\vec{p} \int d\vec{r} \langle \Delta \vec{p_{i}} \rangle_{p,r} = 0
\end{equation}
where the last equality follows from the fact that the collisions conserve
total momentum.
Similarly consider the change in the second moment of the distribution
function
\begin{equation} \label{eqn16}
\Delta \int d\vec{p} \int d\vec{r} f^{(1)} p^{2} = \int d\vec{p}
\int d\vec{r} \Delta f^{(1)} p^{2}.
\end{equation}
In this case two terms remain after integration by parts giving
\begin{equation} \label{eqn17}
\Delta \int d\vec{p} \int d\vec{r} f^{(1)} p^{2} = 
\int d\vec{p} \int d\vec{r} \langle \left( 2\vec{p_{i}}+\Delta \vec{p_{i}}
\right) . \Delta \vec{p_{i}} \rangle_{p,r}.
\end{equation}

Substituting in the momentum evolution equation (\ref{eqn4}) and neglecting
all terms that are first order in momenta or $\xi$ because they
will be zero inside the $\overline{\langle \rangle_{p,r}}$ average
\begin{eqnarray} \label{eqn18}
\overline{\Delta \int d\vec{p} \int d\vec{r} f^{(1)} p^{2}} &=& 
\overline{
\int d\vec{p} \int d\vec{r} \langle 2p_{i}^{\gamma} L_{ik}^{\gamma \delta}
p_{k}^{\delta} + L_{ij}^{\gamma \beta} L_{ik}^{\gamma \delta} p_{j}^{\beta}
p_{k}^{\delta} + M_{ij}^{\gamma} M_{ik}^{\gamma} \xi_{ik} \xi_{ij}
\rangle_{p,r}} \\
\label{eqn19}
&=& \int d\vec{r} \langle mk_{B}T
\left\{ 2 \sum_{\gamma} L_{ii}^{\gamma \gamma} + \sum_{j \beta \gamma}
(L_{ij}^{\beta \gamma})^{2} \right\} + \sum_{j \gamma}
(M_{ij}^{\gamma})^{2} \rangle_{r}.
\end{eqnarray}
Using the expressions (\ref{eqn5}, \ref{eqn6}) this
integral can be evaluated recalling that the distribution function
(\ref{eqn8}) is trivial in position space.
\begin{eqnarray} \label{eqn20}
\lefteqn{\Delta \int d\vec{p} \int d\vec{r} f^{(1)} p^{2} =} \nonumber \\ && 
-2\gamma \Delta t k_{B}T n \left[ w_{D} \right] +
\sigma^{2} \Delta t \left[ w_{R}^{2} \right] n +
\frac{k_{B}T \gamma^{2} (\Delta t)^{2}}{m}
\left\{ 2n \left[ w_{D}^{2} \right] + \frac{ n^{2} \left[ w_{D} \right]^{2}}{d}
\right\}.
\end{eqnarray}
Where d is the number of space dimensions
and the square brackets denote the integral
\begin{equation} \label{eqn21}
\left[ f(\vec{r}) \right] = \int d\vec{r} f(\vec{r}).
\end{equation}
The expression (\ref{eqn20}) must be zero for the one particle
distribution function to remain unchanged
\begin{equation}
mk_{B}T_{eq} = \frac{A_{3}}{A_{1} \left( 2 - A_{1} n\Delta t
\right) -A_{2} \Delta t}
\label{eqn22}
\end{equation}
where
\begin{eqnarray} \label{eqn23}
A_{1} = \frac{\gamma}{md} \left[ w_{D} \right],\;\;\; &
A_{2} = \frac{2\gamma^{2}}{m^{2}d} \left[
w_{D}^{2}\right] ,  \;\;\;&
A_{3} = \frac{\sigma^{2}}{d} \left[ w_{R}^{2} \right]
\end{eqnarray}

We also note that, for the distribution function  (\ref{eqn8})
higher moments are related by
\begin{equation}
\int d\vec{p} f^{(1)} p^{n+2} \propto \int d\vec{p} f^{(1)} p^{n}.
\end{equation}
Therefore, if the constraint (\ref{eqn22}) is satisfied and momentum is
conserved (\ref{eqn15}), all moments of $f^{(1)}$, and therefore
$f^{(1)}$ itself, will remain constant.

We make the following comments on the result (\ref{eqn22}):
\begin{enumerate}
\item For $\Delta t \rightarrow 0$ and $w_{D}=w_{R}^2$
	the formula obtained by Espa\~nol
	and Warren \onlinecite{EW95} is recovered.
\item For a given $(\gamma,\sigma,w_{D},w_{R},n)$
	the measured temperature of the system will
	increase as the time step becomes larger.
\item For a given set of input parameters, the system will become unstable
        for time steps $\Delta t>\Delta t_{c}$ where
	\begin{equation}
	\Delta t_{c} = (2A_{1})/(nA_{1}^{2}+A_{2}).
	\end{equation}
\item Similarly, once a value of $\Delta t$ is chosen, the density
	must not be allowed to exceed a critical density 
	\begin{equation}
	n_{c} = (2A_{1}-A_{2}\Delta t)/(A_{1}^{2}\Delta t)
	\end{equation}
	for a stable simulation.
\item The choice of a small value of $\gamma$ will decrease the effect
	of a finite $\Delta t$.
\end{enumerate}

Simulation results, shown in Figure 1, show that equation (\ref{eqn22})
gives a good prediction of the dependence of the temperature on the time
step for several densities
and time steps.  The simulations were run in two dimensions with 
periodic boundary conditions. The system size was 100 units and the 
interaction was range was 4 units. Averages were taken
over 9 runs each of duration 1000. Error bars are of the 
order of the size of the points in Figure 1.
The small deviations between the analytic and numerical results
arise because correlations between particles modify the Gibb's
distribution (\ref{eqn8}) for finite $\Delta t$.  However, the result
(\ref{eqn22}) is a good prediction of the temperature of the system
over a wide range of system parameters relevant to numerical simulations.

To conclude, we have demonstrated that, for a DPD simulation of an
ideal fluid, the equilibrium temperature of the
system depends strongly on the time step. This implies that 
caution is necessary if DPD is used
to probe equilibrium thermodynamic properties.

\section*{Acknowledgements}

We thank Antoine Schlijper for introducing us to DPD, Peter
Coveney, Alexander Wagner, Patrick Warren
for helpful comments and Matthieu Ernst for pointing out
algebraic errors.  We acknowledge support from
the EPSRC, UK and a CASE studentship from Unilever Research.

\begin{figure}
\vspace{1.5in}
\inseps{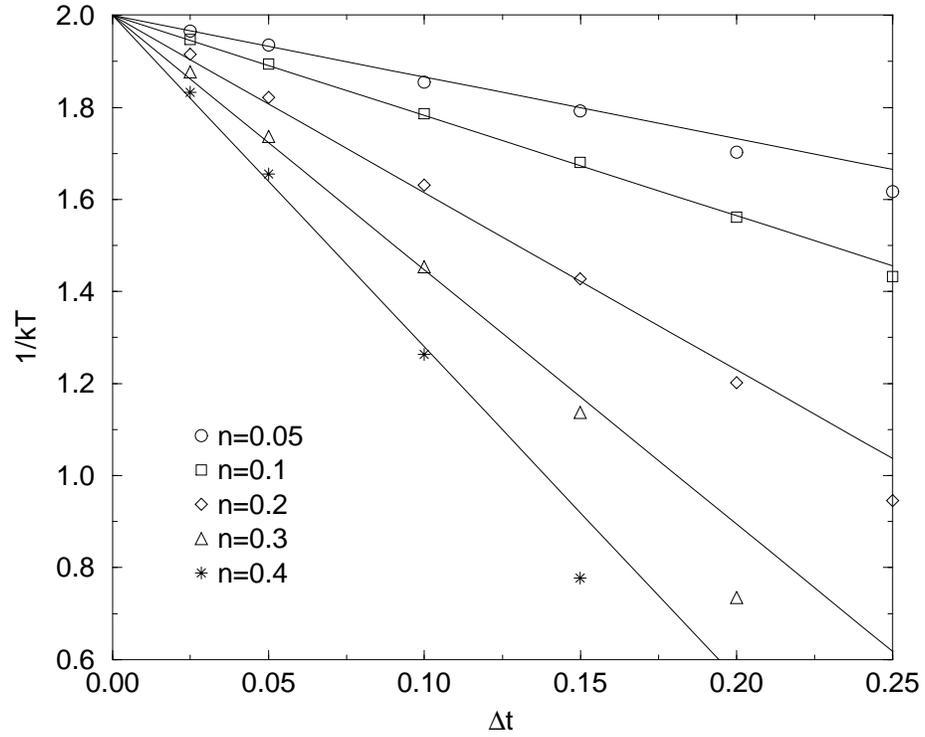}{0.7}
\caption{Dependence of the inverse temperature, $1/T$, on the time step
$\Delta t$ for different values of the density $n$ for DPD simulations
of an ideal fluid. The system parameters were
$\gamma=1$, $\sigma=1$. The lines correspond to the
predictions of equation (22).}
\end{figure}

\end{document}